\documentclass[11pt,a4paper]{article}
\usepackage{jheppub}
\usepackage{epsf}
\usepackage{amsmath}
\usepackage{amssymb}
\usepackage{amsfonts}
\usepackage{bbold}
\usepackage{graphicx}
\usepackage{graphics}
\usepackage{epstopdf}
\usepackage{url}

\newcommand{\beq}{\begin{equation}}
\newcommand{\eeq}{\end{equation}} 
\newcommand{\beqa}{\begin{eqnarray}} 
\newcommand{\eeqa}{\end{eqnarray}}

\def\bc{\begin{center}}
\def\ec{\end{center}}
\def\bi{\begin{itemize}}
\def\ei{\end{itemize}}
\def\be{\begin{equation}}
 \def\ee{\end{equation}}
\def\ben{\begin{equation*}}
 \def\een{\end{equation*}}
 \def\bea{\begin{eqnarray}}
 \def\eea{\end{eqnarray}}
 \def\bean{\begin{eqnarray*}}
 \def\eean{\end{eqnarray*}}
\newcommand{\ie}{{\em i.e.}}

\newcommand{\morder}[1]{{\cal O}\left(#1 \right)}
\newcommand{\eq}[1]{(\ref{#1})}
\newcommand{\ed}{\end{document}}

\newcommand{\com}[2]{\left[{#1},{#2}\right]}

\newcommand{\pvec}{{\boldsymbol p}}
\newcommand{\ellvec}{{\boldsymbol \ell}}
\newcommand{\kvec}{{\boldsymbol k}}
\newcommand{\Kvec}{{\boldsymbol K}}
\newcommand{\qvec}{{\boldsymbol q}}

\newcommand{\epsvec}{{\boldsymbol \varepsilon}}

\newcommand{\tf}{t_{\mathrm{f}}}

\newcommand{\jpsi}{{\mathrm J}/\psi}

\newcommand{\xh}{x_{\mathrm{h}}}

\newcommand{\dd}{{\rm d}}
  
\newcommand{\lsim}{\lesssim} \newcommand{\gsim}{\gtrsim}

 \def\esim{\,\mathrel{\rlap{\lower0.2em\hbox{$-$}}\raise0.15em\hbox{\footnotesize $\hskip0.04em\sim$}}\,}
 \def\gsim{\mathrel{\rlap{\lower0.2em\hbox{$\sim$}}\raise0.2em\hbox{$>$}}}
 \def\ksim{\mathrel{\rlap{\lower0.2em\hbox{$\sim$}}\raise0.2em\hbox{$<$}}}

\title{On the process-dependence of coherent medium-induced gluon radiation}

\author[a]{St\'ephane Peign\'e,}
\author[a,b]{Rodion Kolevatov}

\affiliation[a]{SUBATECH, UMR 6457, Universit\'e de Nantes, Ecole des Mines de Nantes, IN2P3/CNRS \\ 4 rue Alfred Kastler, 44307 Nantes cedex 3, France}
\affiliation[b]{Department of High Energy Physics, Saint-Petersburg State University\\ Ulyanovskaya 1, 198504, Saint-Petersburg, Russia}

\emailAdd{peigne@subatech.in2p3.fr}
\emailAdd{kolevato@subatech.in2p3.fr}

\abstract{
Considering forward dijet production in the $q\to qg$ partonic process, we derive the spectrum of accompanying soft gluon radiation induced by rescatterings in a nuclear target. The spectrum is obtained to logarithmic accuracy for an arbitrary energy sharing between the final quark and gluon, and for final transverse momenta as well as momentum imbalance being large as compared to transverse momentum nuclear broadening. In the case of equal energy sharing and for approximately back-to-back quark and gluon transverse momenta, we reproduce a previous result of Liou and Mueller. Interpreting our result, we conjecture a simple formula for the medium-induced radiation spectrum associated to hard forward $1 \to n$ processes, which we explicitly check in the case of the $g \to gg$ process.}


\begin{document}

\maketitle
\setcounter{footnote}{0}
\renewcommand{\thefootnote}{\arabic{footnote}} 	

\section{Introduction and summary}

A few years ago it was emphasized that the medium-induced radiative energy loss $\Delta E$ of a high-energy gluon crossing a nuclear medium and being scattered to small angle (in the medium rest frame) is proportional to the gluon energy $E$~\cite{Arleo:2010rb}. The behavior $\Delta E \propto E$ arises from soft gluon radiation with formation time $\tf$ scaling as $E$, \ie, being {\it fully coherent} over the size $L$ of the medium ($\tf \gg L$ at large $E$). As discussed in Ref.~\cite{Arleo:2010rb}, coherent radiative energy loss arises from the interference between emission amplitudes off the incoming and outgoing particles, and is thus expected in all situations where the hard partonic process is effectively equivalent to the forward scattering of an incoming parton to an outgoing compact {\it colored} system of partons. Coherent energy loss should play an important role in the high-energy {\it hadro}production of hadrons, but should be absent in (inclusive) Drell-Yan production, as well as in hadron 
photoproduction. In the case of $\jpsi$ hadroproduction at low $p_\perp \lesssim M_{\jpsi}$, viewed in the target rest frame as the scattering of an incoming gluon to an outgoing {\it color octet} compact $c \bar{c}$ pair, such a coherent, medium-induced energy loss was shown to provide a successful description of $\jpsi$ nuclear suppression in proton-nucleus (p--A) as compared to proton-proton (p--p) collisions, from fixed-target (SPS, HERA, FNAL) to collider (RHIC, LHC) energies \cite{Arleo:2012hn,Arleo:2012rs,Arleo:2013zua}. 

Before studying the possible effect of coherent energy loss on other observables, one should first consider the question of the process dependence of the medium-induced coherent radiation spectrum $\omega \dd I/\dd \omega$. Recent studies~\cite{Liou:2014rha,Peigne:2014uha} started to address this question. The radiation spectra associated to $1\to 1$ \cite{Peigne:2014uha} and $1\to 2$ \cite{Liou:2014rha} forward scattering processes are found to be proportional to the same logarithm of the kinematical parameters, but to possibly differ by an overall factor. For instance, Liou and Mueller showed that the $q\to qg$ and $g \to q \bar{q}$ processes, in the kinematics where the outgoing jets have identical longitudinal momenta and nearly back-to-back transverse momenta, lead to the same medium-induced radiation spectrum up to a surprising factor $4/5$ \cite{Liou:2014rha}. 

In the present study, we derive the coherent radiation spectrum associated to the $q\to qg$ process already studied in \cite{Liou:2014rha}, but using a different setup. First, we consider the outgoing gluon and quark to carry the fractions $\xh$ and $(1-\xh)$ of the incoming (light-cone) longitudinal momentum $p^+$ ($\xh = 1/2$ was chosen in \cite{Liou:2014rha}). Second, not only the final gluon and quark transverse momenta $\vec{K}_{1\perp} \equiv \Kvec_1$ and $\vec{K}_{2\perp} \equiv \Kvec_2$, {\it but also their momentum imbalance $\Kvec_1 + \Kvec_2 \equiv \qvec$}, are chosen to be hard compared to the transverse momentum broadening $\Delta q_\perp$ across the medium, $|\Kvec_1|, |\Kvec_2|, |\qvec| \gg \Delta q_\perp$. Within this setup, the radiation spectrum associated to $q\to qg$ is derived in an opacity expansion (as in \cite{Peigne:2014uha} for $1 \to 1$ forward processes), and in the large $N_c$ limit (as in~\cite{Liou:2014rha}). The resulting radiation spectrum is proportional to the same leading 
logarithm as in Refs.~\cite{Liou:2014rha,Peigne:2014uha}, but with a prefactor depending on the hard $q\to qg$ process through the kinematical variables $\xh$, $\qvec$, $\Kvec_1$. This prefactor is simply interpreted as the probability for the $qg$ pair to be produced in the ${\bf \bar{6} \oplus 15}$ subspace of all possible $qg$ color states. For $|\qvec| \ll |\Kvec_1|$ and $\xh = 1/2$ we recover the factor $4/5$ found in \cite{Liou:2014rha}. We conjecture the simple formula \eq{conjecture} for the medium-induced radiation spectrum associated to hard forward $1 \to n$ processes. The conjecture is explicitly verified in the case of the $g \to gg$ process, where in the particular limit $|\qvec| \ll |\Kvec_1|$ and $\xh = 1/2$ we find the overall factor $5/3$, instead of $4/5$ for the  
$q\to qg$ process. 

In section~\ref{sec:gtog} we review the theoretical setup and the results of Refs.~\cite{Arleo:2010rb,Peigne:2014uha} for $1\to 1$ forward processes, and give a physical interpretation of the main features of the medium-induced coherent radiation spectrum. The setup and calculation are generalized to the hard $q \to qg$ process in section~\ref{sec:q2qg}.


\section{Forward single jet production}
\label{sec:gtog}

\subsection{Review of previous studies}

\begin{figure}[t]
\centering
\includegraphics[width=13cm]{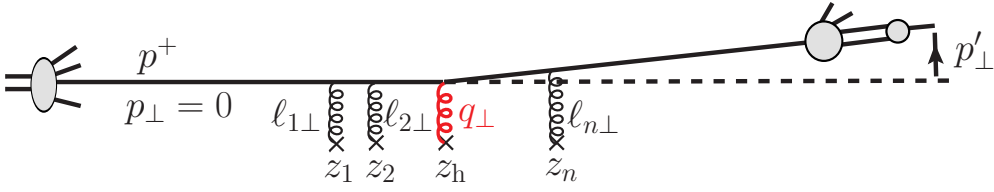}
\caption{Setup for $1 \to 1$ forward production used in \cite{Arleo:2010rb,Peigne:2014uha}, as viewed in the nuclear target rest frame. The solid line denotes the energetic parton (gluon or quark). The hard process is modeled by a transverse momentum exchange $q_\perp$ (in red) occurring at the longitudinal position $z_{\rm h}$, and supplemented by soft rescatterings $\ell_{i\perp} \ll q_\perp$ (with also $\ell_\perp = |\sum \ellvec_i| \ll q_\perp$) occurring at longitudinal positions $z_i$.}
\label{fig:hard-elastic}
\end{figure}

Consider a massless parton of large momentum $p=(p^+,0,\vec{0}_\perp)$ with $p^+\equiv 2E$,\footnote{We use light-cone variables, $p=(p^+,p^-,\pvec)$, with $p^{\pm} = p^0 \pm p^z$ and $\pvec \equiv \vec{p}_\perp$.} prepared in the far past and traversing some nuclear medium, see Fig.~\ref{fig:hard-elastic}. The final energetic `jet' is `tagged' with a transverse momentum $\pvec'$ much larger than the nuclear transverse broadening $\ellvec = \sum \ellvec_i$ acquired through multiple {\em soft} scattering. As a consequence $\pvec'$ must arise dominantly from a {\em single hard} scattering $\qvec \simeq \pvec'$, with $|\qvec| \gg |\ellvec|$. We focus on the $p^+ \to \infty$ limit at fixed transverse momentum (small angle scattering). The fast parton is also assumed to scatter with a negligible longitudinal momentum transfer to the medium, which allows one to neglect the recoil of the target partons. This setup is used in \cite{Arleo:2010rb,Peigne:2014uha} to derive the medium-induced coherent radiation associated to `forward single 
jet' production (\ie, $1 \to 1$ forward production), which we briefly review below. In the following, the radiated gluon momentum is denoted by $k=(k^+,\kvec^2/k^+,\kvec \equiv \vec{k}_\perp)$, and we focus on soft ($x \equiv k^+/p^+ \ll 1$) and small angle ($|\kvec| \ll k^+$) radiation (hence $k^{+} \equiv \omega + k^z \simeq 2 \omega$). As we will see, the main features of coherent radiation induced by $1 \to 1$ processes also arise for the $1 \to 2$ process ($q \to qg$) studied in section \ref{sec:q2qg}. This is because coherent radiation, in the limit considered in section \ref{sec:q2qg}, effectively sees the `dijet' $qg$ final state as a pointlike object.
 
In Ref.~\cite{Arleo:2010rb} the coherent radiation spectrum $\dd I/\dd x$ associated to the $g \to g$ process (with the final `gluon' being a compact color octet $Q \bar{Q}$ pair of mass $M$) was derived by modeling the transverse momentum broadening $\Delta q_\perp$ across the medium by a {\it single} rescattering $\ell_\perp$, and identifying $\ell_\perp^2 = \Delta q_\perp^2(L) = \hat{q} L$, with $\hat{q}$ the transport coefficient. The obtained result, 
\be
\label{spec-alln-appr}
\left. x \frac{\dd I}{\dd x} \right|_{g \to g}  = N_c \, \frac{\alpha_s}{\pi} \, \log{\left( 1 + \frac{\Delta q_\perp^2(L)}{x^2 M_\perp^2} \right) } \, ,
\ee
where $M_\perp^2  \equiv M^2 + p'^{\,2}_{\perp} \simeq M^2 + q_\perp^2$, was confirmed in \cite{Peigne:2014uha} in a theoretical setup using the opacity expansion \cite{Gyulassy:2000er}, allowing one to consider an arbitrary number $n$ of soft rescatterings in the medium. 

In the present paper we focus on the case of massless particles and, similarly to Ref.~\cite{Liou:2014rha}, on the small-$x$ region where the spectrum is logarithmically enhanced. We thus rewrite \eq{spec-alln-appr} as 
\be
\label{spec-alln-LL-gg}
\left. x \frac{\dd I}{\dd x} \right|_{g \to g}  = N_c \, \frac{\alpha_s}{\pi}  \left[ \log{\left(\frac{\Delta q_\perp^2(L)}{x^2 q_\perp^2} \right) }  + \morder{1} \right] \, .
\ee
We stress that the latter expression holds when not only the argument of the logarithm, but the logarithm itself is much larger than unity,  \ie, to {\it logarithmic accuracy}, which will be implicit throughout our study.

\begin{figure}[t]
\centering
\includegraphics[width=12cm]{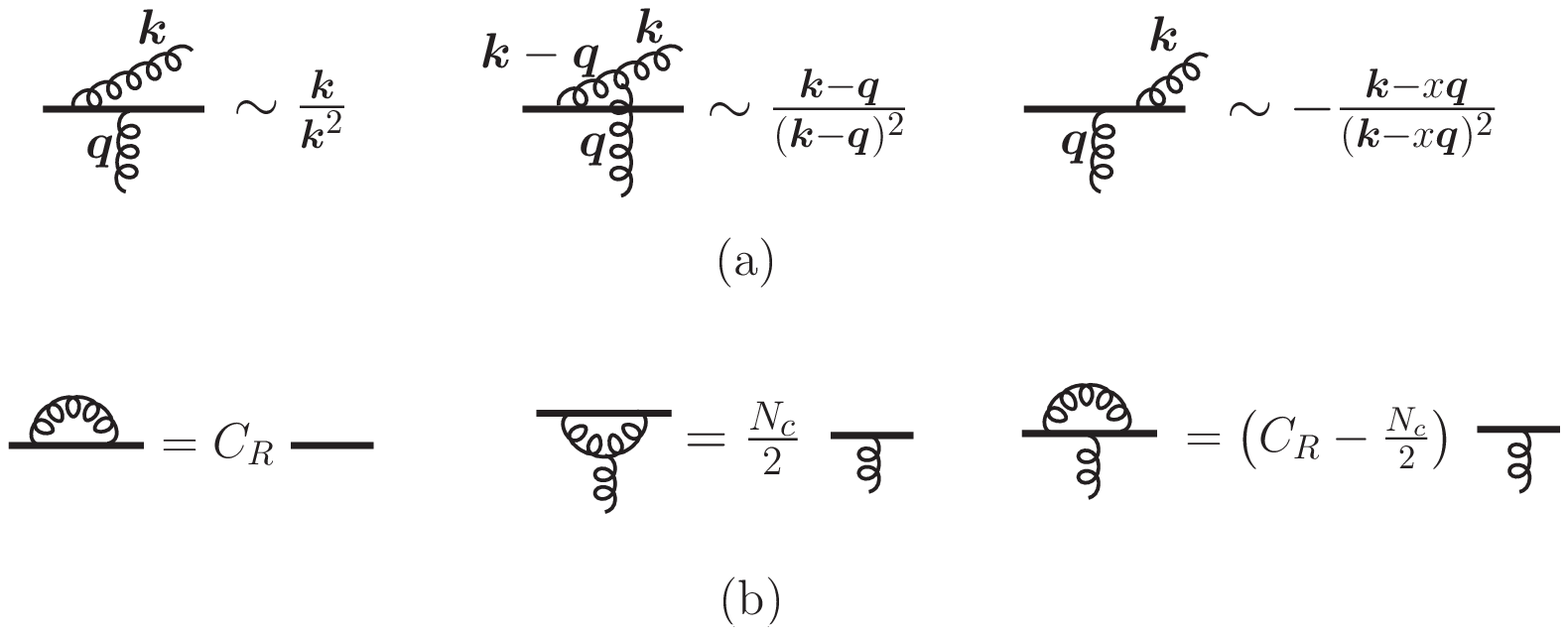}
\caption{Pictorial rules for (a) emission vertices and (b) color factors. The energetic parton of color charge $C_R$ is denoted by the solid line ($C_R = C_A =N_c$ for a gluon and $C_R = C_F= (N_c^2 -1)/(2N_c)$ for a quark). For the pictorial representation of color factors, see for instance Ref.~\cite{Dokshitzer:1995fv}.}
\label{fig:pictorial-rules}
\end{figure}

It is useful to recall the basic steps leading to \eq{spec-alln-LL-gg}. It was shown in \cite{Peigne:2014uha} that the spectrum at order $n$ in the opacity expansion is given by
\be
x \frac{\dd I^{(n)}}{\dd x}  =
\frac{\alpha_s}{\pi^2} \int \dd^2 \kvec \left[ \prod_{i=1}^{n} \int \frac{\dd z_i}{C_R \lambda_R} \int \dd^2 \ellvec_i \, V(\ellvec_i) \right]  \, \frac{\sum \: \mbox{\raisebox{-7mm}{\hskip 0mm \hbox{\epsfxsize=4.7cm\epsfbox{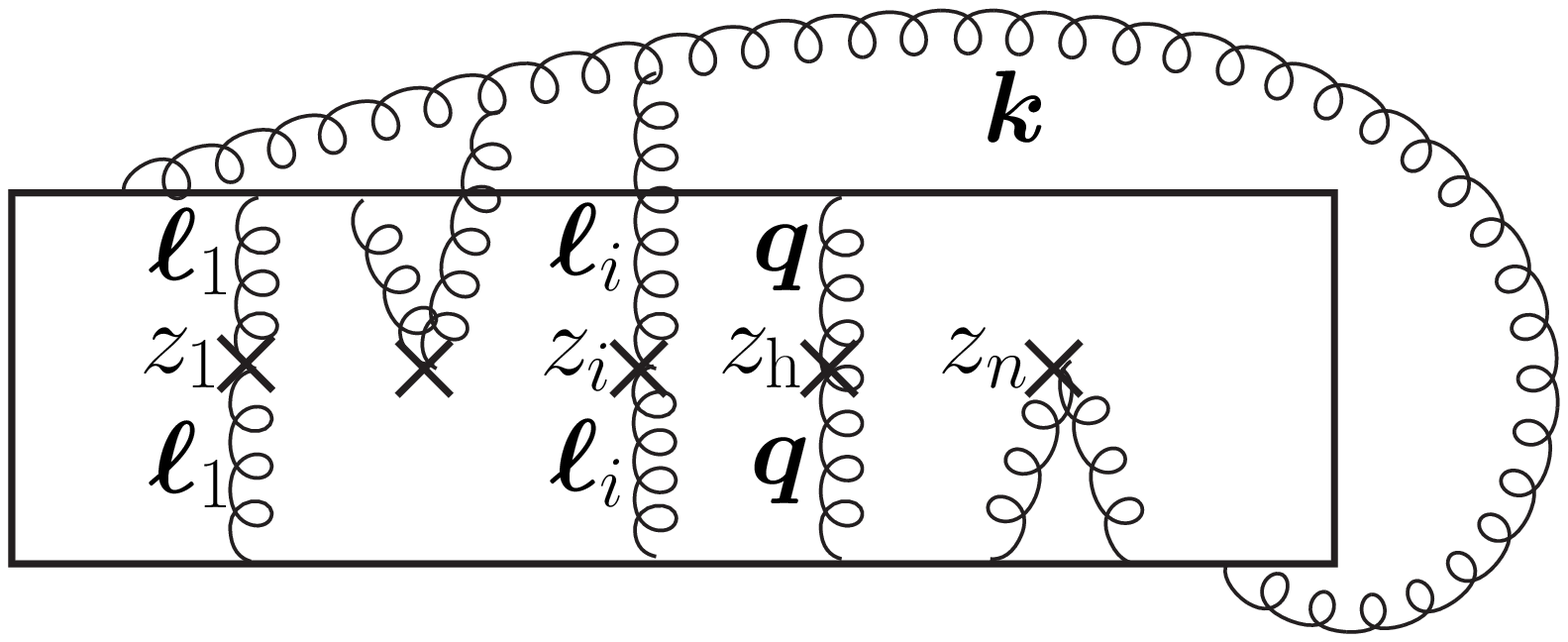}}}}}{ \mbox{\raisebox{-7mm}{\hskip 0mm \hbox{\epsfxsize=1.7cm\epsfbox{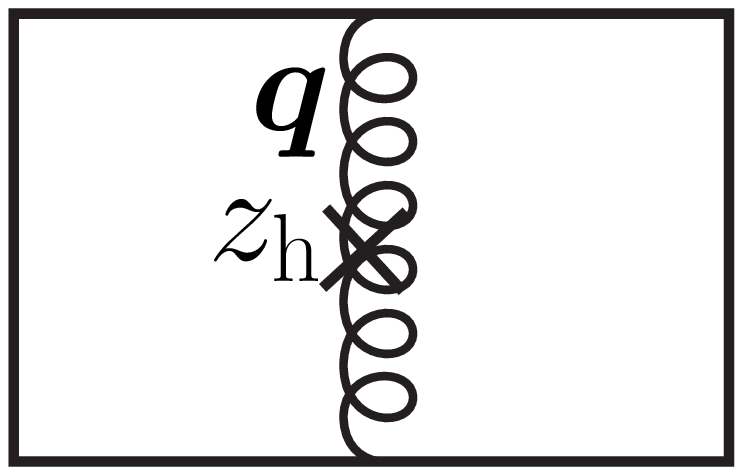}}}}}
\hskip 0mm 
\label{spec-order-n} 
\ee
where the diagrams appearing in the numerator and denominator are evaluated using the pictorial rules defined in Fig.~\ref{fig:pictorial-rules}. The upper (lower) part of each diagram appearing in the numerator of  \eq{spec-order-n} corresponds to a contribution to the emission amplitude (conjugate amplitude) of the soft gluon $\kvec$ induced by the rescatterings $\ellvec_i$. The diagram in the denominator stands for the hard process `cross section' (which here is a single color factor, other factors cancelling between numerator and denominator). The quantity $\lambda_R$ is the  elastic mean free path of the fast parton of color charge $C_R$ (note that the product $C_R \lambda_R=C_F \lambda_q = N_c \lambda_g$ is independent of the parton type), and an average over soft transfers $\ellvec_i$ is performed using the screened Coulomb potential $V(\ellvec_i) = \mu^2/[\pi (\ellvec_i^2 + \mu^2)^2]$. The latter provides the typical magnitude of soft transfers, $|\ellvec_i| \sim \mu  \ll |\qvec|$, with $\mu$ being 
the inverse screening length of the medium. 

We stress that the evaluation of \eq{spec-order-n} in \cite{Peigne:2014uha} is done in the {\it coherent} limit $\tf \gg L$ and assuming $k_\perp \ll q_\perp$, which leads to important simplifications:
\bi 
\item[(i)] Diagrams where the time $t$ associated to the soft emission vertex is in between two rescatterings, $z_i < t < z_{i+1}$, are negligible. 
\item[(ii)] Diagrams where the hard gluon $\qvec$ couples to the soft radiated gluon $\kvec$ are negligible. (The rescattering gluons $\ellvec_i$ can couple to both the energetic parton and the soft gluon, including virtual contributions where two gluon lines $\ellvec_i$ and $-\ellvec_i$ are transferred in either the amplitude or conjugate amplitude.)
\item[(iii)] At each order in opacity, the contribution to \eq{spec-order-n} of purely initial state radiation cancels out. (The same holds for purely final state radiation.) Only interference diagrams remain, like the generic diagram drawn in numerator of \eq{spec-order-n}.
\ei

At first order in opacity we find \cite{Peigne:2014uha}
\be
\label{spectrum-n1}
x \frac{\dd I^{(1)}}{\dd x}  = (2 C_R - N_c) \, \frac{\alpha_s}{\pi^2} \, \frac{L}{\lambda_g}  \int \dd^2 \kvec  \int \dd^2 \ellvec_1 \, V(\ellvec_1) \, \left[ \frac{\kvec -  \ellvec_1}{(\kvec - \ellvec_1)^{2}} -  \frac{\kvec}{\kvec^{2}} \right] \cdot \frac{-(\kvec - x\qvec)}{(\kvec - x\qvec)^{2}} \, ,
\ee
which can be interpreted as the interference between the wavefunction of the final parton-gluon fluctuation (the last factor in the integrand of \eq{spectrum-n1}), and the incoming `medium-induced wavefunction'  (factor in between brackets). For the purpose of the present study, it is sufficient to observe that when $x |\qvec| \ll |\ellvec_1| \sim \mu$, the spectrum arises from the logarithmic $k_\perp$-domain $x |\qvec| \ll |\kvec| \ll \mu$, leading to
\be
\label{specn1-LL}
x \frac{\dd I^{(1)}}{\dd x}  = (2 C_R - N_c ) \, \frac{\alpha_s}{\pi} \, \frac{L}{\lambda_g} \, \log{\left(\frac{\mu^2}{x^2 \qvec^2} \right) } \, .
\ee

The spectrum at all orders in opacity derived in \cite{Peigne:2014uha} can be formally obtained from \eq{specn1-LL} by shifting the rescattering probability $L/\lambda_g$ by unity, and the broadening in a {\it single} scattering $\mu^2$ by the broadening in {\it multiple} scattering $\Delta q_\perp^2(L) \sim \mu^2 L/\lambda_g \equiv \hat{q} L$. It reads
\be
\label{spec-alln-LL}
x \frac{\dd I}{\dd x} = \sum_{n=1}^{\infty} x \frac{\dd I^{(n)}}{\dd x} = (2 C_R - N_c ) \, \frac{\alpha_s}{\pi} \,  \log{\left(\frac{\Delta q_\perp^2(L)}{x^2 \qvec^2} \right) } \, ,
\ee
which in the case of an incoming gluon ($C_R = N_c$) yields the result \eq{spec-alln-LL-gg}. 
 
\subsection{Physical interpretation} 
\label{interpret-11}
 
\subsubsection*{logarithmic range}
 
At small $x \ll \Delta q_\perp(L)/|\qvec|$ and to logarithmic accuracy, the spectrum \eq{spec-alln-LL} arises from the region
\be
\label{krange-zeromass}
x |\qvec| \ll |\kvec| \ll \Delta q_\perp(L) \, ,
\ee
which has a simple physical interpretation.

First, the leftmost inequality is equivalent to saying that at the time $\tf \sim \omega/k_\perp^2$ of its emission, the soft gluon does {\it not} probe the relative displacement $\Delta \vec{r}$ of the {\it core} charge compared to the case of unperturbed (vacuum) propagation. Indeed, denoting $v_1$ ($v_2$) the velocity of the incoming (outgoing) energetic charge, the latter statement reads~\cite{Dokshitzer:1995if}
\be
\label{interpret-krange-zeromass}
1/\omega \gg \Delta r_\parallel = |v_{2\parallel}- v_{1\parallel}|\,\tf \sim \frac{q_\perp^2}{E^2} \frac{\omega}{k_\perp^2} \ \ \ {\it and } \ \ \ 1/k_\perp \gg \Delta r_\perp = v_{2\perp}\,\tf \sim \frac{q_\perp}{E} \frac{\omega}{k_\perp^2} \, , 
\ee
which is equivalent to the single condition $x q_\perp \ll k_\perp$. Under such a constraint, no radiation would occur in QED, where only the photon field components which can `see' the deviation of the parent charge can be released as true radiation. In the present QCD situation, the inequality $x q_\perp \ll k_\perp$ thus means that only the purely non-abelian part of radiation contributes to the spectrum. But for this part of radiation to actually contribute to the {\it medium-induced} spectrum, the soft gluon should probe the transverse displacement $\Delta r_\perp^{\mathrm g}$ of the core charge {\it proper gluon field} induced by rescatterings. This implies $1/k_\perp \ll \Delta r_\perp^{\rm g} \sim (\ell_\perp/\omega)\,\tf$, where $\ell_\perp/\omega$ is the deviation angle of the incoming gluon proper field in the medium. This leads to the second inequality of \eq{krange-zeromass}. In summary, the range \eq{krange-zeromass} can be interpreted as the dominant $k_\perp$-region for {\it purely non-
abelian}, {\it medium-induced} radiation.

 
\subsubsection*{color factor} 
 
The color factor associated to the coherent radiation spectrum \eq{spec-alln-LL} can also be simply understood. For a general $1 \to 1$ process with incoming and outgoing particles in color representations $R$ and $R'$, respectively, the color factor is $2 T_R^a T_{R'}^a$, as can be trivially checked from the structure of the interference terms giving rise to \eq{spec-alln-LL}. Using the identity
\be
\label{color-rule}
2 T_R^a T_{R'}^a = (T_R^a)^2 + (T_{R'}^a)^2  - (T_{R}^a - T_{R'}^a)^2 = C_R + C_{R'} - C_{t} \, ,
\ee
where $C_{t}$ is the color charge exchanged in the $t$-channel of the hard process, we recover the factor $2 C_R-N_c$ in the case \eq{spec-alln-LL} of asymptotic parton scattering.\footnote{\label{foot:quark-loss}For $q \to q$ scattering, this factor reads $2 C_F-N_c =-1/N_c$, and the medium-induced radiation spectrum associated to $q \to q$ is thus suppressed in the large $N_c$ limit.} For the processes $q \to g$ and $g \to q$ mediated by $t$-channel color triplet exchange, the color factor reads $C_F + N_c -C_F =N_c$, as found in \cite{Peigne:2014uha}.

\section{Forward $q \to qg$ production}
\label{sec:q2qg}

We now consider a simple generalization of section \ref{sec:gtog}, by replacing the $1 \to 1$ forward hard process by the $q \to qg$ process, and derive the associated medium-induced soft radiation spectrum.

\subsection{Model for $q \to qg$ hard process}

\begin{figure}[t]
\centering
\includegraphics[width=12cm]{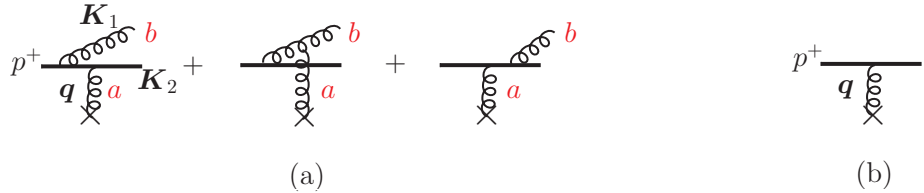}
\caption{(a) Amplitude for hard $q \to qg$ production process. (b) Elastic amplitude ${\cal M}_{\rm el}$.}
\label{fig:GBamplitude}
\end{figure}

The $q \to qg$ production amplitude is depicted in Fig.~\ref{fig:GBamplitude}a, where the final gluon and quark have transverse momenta $\Kvec_1 \equiv \vec{K}_{1\perp}$ and $\Kvec_2 \equiv \vec{K}_{2\perp}$, and light-cone longitudinal momentum fractions $\xh \equiv K_1^+/p^+$ and $1-\xh \equiv K_2^+/p^+$, respectively. We consider the $p^+ \to \infty$ limit at fixed and {\it finite} $\xh \sim \morder{1}$. The amplitude of Fig.~\ref{fig:GBamplitude}a is conveniently derived in a light-cone formalism and in light-cone $A^+ =0$ gauge. 

In {\it scalar} QCD, we find
\be
\label{hard-scalarQCD}
{\cal M}_{\rm hard} =  \hat{\cal M}_{\rm el} \cdot 2g\,(1-\xh) \cdot \left[  T^aT^b \, \frac{\Kvec_1}{\Kvec_1^{2}} + \com{T^b}{T^a} \, \frac{\Kvec_1 - \qvec}{(\Kvec_1 - \qvec)^{2}} - T^bT^a \, \frac{\Kvec_1 - \xh \qvec}{(\Kvec_1 - \xh \qvec)^{2}} \right] \cdot \epsvec_1 \, ,
\ee
where $\hat{\cal M}_{\rm el}$ denotes the Lorentz part (\ie, without color factor) of the elastic amplitude of Fig.~\ref{fig:GBamplitude}b , and $\epsvec_1 \equiv \vec{\varepsilon}_{1\perp}$ the final gluon physical polarization. Since $\epsvec_1$ formally disappears after squaring the amplitude and summing over the two physical polarization states, it can be dropped in \eq{hard-scalarQCD}. 

In QCD, the spinor structure makes the amplitude of Fig.~\ref{fig:GBamplitude}a slightly more complicated than \eq{hard-scalarQCD}.\footnote{The QCD calculation can be done using light-cone helicity spinors~\cite{Lepage:1980fj}.}  However, after squaring, summing over polarization states and over color, the result for $|{\cal M}_{\rm hard}|^2$ in QCD is the same (for a given quark light-cone helicity) {\it as if} ${\cal M}_{\rm hard}$ were given by the scalar QCD expression \eq{hard-scalarQCD}, up to the replacement of the overall factor
\be
1-\xh \ \ ({\rm scalar \ QCD}) \  \to \  \sqrt{\frac{1+(1-\xh)^2}{2}}  \ \ ({\rm QCD}) \, .
\ee
Moreover, specific contributions to $|{\cal M}_{\rm hard}|^2$ corresponding to the interference of different graphs of Fig.~\ref{fig:GBamplitude}a (including initial and final state radiation) are reproduced one by one with this replacement. We specially emphasize this fact since only part of these contributions enter the calculation of the induced soft radiation spectrum (see section~\ref{sec:indspec}). 

The overall factor is irrelevant for our purpose, since it will cancel between numerator and denominator in the induced soft radiation spectrum \eq{specn1-qg-0}. We can thus use (in either spinor or scalar QCD):
\be
\label{GB}
{\cal M}_{\rm hard} \  \propto  \ \ T^aT^b \, \frac{\Kvec_1}{\Kvec_1^{2}} + \com{T^b}{T^a} \, \frac{\Kvec_1 - \qvec}{(\Kvec_1 - \qvec)^{2}} - T^bT^a \, \frac{\Kvec_1 - \xh \qvec}{(\Kvec_1 - \xh \qvec)^{2}}  \ \, .
\ee
We stress that this expression, derived long ago by Gunion and Bertsch~\cite{Gunion:1981qs},\footnote{In Ref.~\cite{Gunion:1981qs}, the scalar QCD expression \eq{hard-scalarQCD} is given, but used only in the limit $\xh \to 0$, where the scalar QCD and spinor QCD expressions of $|{\cal M}_{\rm hard}|^2$ coincide (after summing over gluon polarization states).} holds for any {\it finite} $\xh$ in the $p^+ \to \infty$ limit.
 
The amplitude \eq{GB} will be our model for the hard process. In addition to $\xh \sim \morder{1}$,  we choose (as in Ref.~\cite{Liou:2014rha}) $\Kvec_1$ and $\Kvec_2$ to be much larger than the nuclear broadening $\Delta q_\perp$.  However, in view of applying the opacity expansion as in the $1\to 1$ case studied in section \ref{sec:gtog}, we also choose the dijet {\it momentum imbalance} $\qvec = \Kvec_1 + \Kvec_2$ to satisfy $|\qvec| \gg \Delta q_\perp$. As a consequence, the dijet imbalance is provided by a single hard exchange $\qvec$ and negligibly affected by soft rescatterings in the medium. In summary we consider the $q \to qg$ process in the kinematics
\be
\xh \sim \morder{1}  \hskip 5mm {\rm and} \hskip 5mm |\Kvec_1|,\, |\Kvec_2|,\,  |\qvec| \, \gg \, \Delta q_\perp \, .
\ee

\subsection{Medium-induced coherent radiation spectrum} \label{sec:indspec}

The medium-induced radiation spectrum associated to $q \to qg$ is derived in the {\it soft} radiation limit defined by
\be
x  \equiv \frac{k^+}{p^+}  \ll 1 \hskip 5mm {\rm and} \hskip 5mm k_\perp \, \ll \, |\Kvec_1|,\,  |\Kvec_2|, \,  |\qvec|  \, .
\ee

\begin{figure}[t]
\centering
\includegraphics[width=13.2cm]{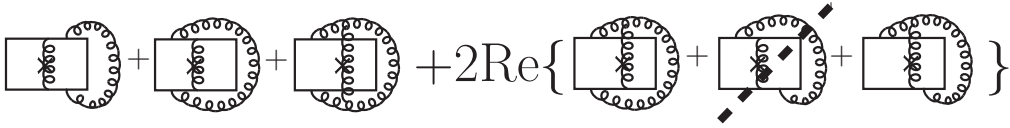}
\caption{Set $H$ of diagrams corresponding to the $q \to qg$ hard process, appearing in the denominator of the radiation spectrum \eq{specn1-qg-0}. The barred diagram is suppressed in the large $N_c$ limit.}
\label{fig:den}
\end{figure}

\begin{figure}[t]
\centering
\includegraphics[width=14cm]{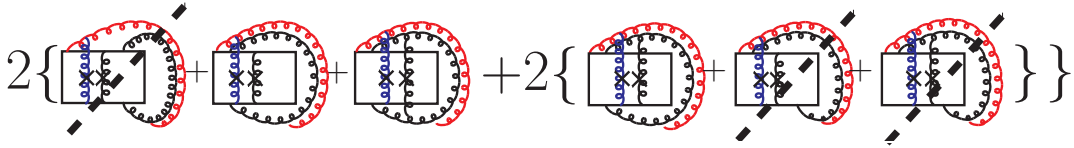}
\caption{Set $C$ of diagrams (numerator of \eq{specn1-qg-0}) for soft gluon emission (in red) induced by a single in-medium rescattering (in blue). Among all possible attachments of the rescattering gluon to the hard quark and gluon lines (including virtual contributions where the rescattering gluon couples to two lines in either the amplitude or conjugate amplitude), only one is drawn. Diagrams which are suppressed at large $N_c$ are barred.}
\label{fig:num}
\end{figure}

The calculation is greatly simplified by observing that the hard process structure \eq{GB} is given by the pictorial rules of Fig.~\ref{fig:pictorial-rules}. It is then straightforward to show that the radiation spectrum associated to $q \to qg$ is given, at order $n$ in opacity, by the expression \eq{spec-order-n} with the $1 \to 1$ replaced by the $q \to qg$ hard process. For instance, at first order in opacity,
\be
\label{specn1-qg-0}
\left. x \frac{\dd I^{(1)}}{\dd x} \right|_{q \to qg}  =
\frac{\alpha_s}{\pi^2} \int \dd^2 \kvec \int \frac{\dd z_1}{N_c \lambda_g} \int \dd^2 \ellvec_1 \, V(\ellvec_1) \, \frac{C}{H} \, , 
\ee
where $H$ and $C$ are given by the diagrams of Figs.~\ref{fig:den} and \ref{fig:num}, respectively. This calls for several comments:
\bi 
\item[(i)] As a simplifying assumption, we choose  the gluon formation time $\tf$ to be large not only compared to $L$ (as in section \ref{sec:gtog}) but also compared to the hard process production time $t_{\rm hard}$,
\be
\label{tf-assumption}
\tf \sim \frac{k^+}{k_\perp^2} \gg t_{\rm hard} \sim \frac{p^+}{K_{1\perp}^2} \gg L \, .
\ee
In this limit, the dominant diagrams are those (as in the set $C$ of Fig.~\ref{fig:num}) where the soft gluon emission vertex is either long before or long after the interaction vertices of the hard process shown in Fig.~\ref{fig:den}. The assumption $\tf \gg t_{\rm hard}$ constrains the range of validity of our final result \eq{qqg-spec} to the domain \eq{validity-range}, but still allows for accessing the main features of the induced spectrum.
\item[(ii)] As in the case of $1 \to 1$ forward processes, diagrams corresponding to purely initial-state or purely final-state radiation cancel out in the medium-induced spectrum. Only interference diagrams remain, where the soft gluon is emitted {\it before} the hard process in the amplitude, and {\it after} in the conjugate amplitude (as in Fig.~\ref{fig:num}). 
\item[(iii)] We work in the large $N_c$ limit \cite{tHooft:1973jz}. In this limit, the interference diagrams where the soft gluon connects to the final quark line are all suppressed, and thus not drawn in Fig.~\ref{fig:num}. Among the diagrams where the soft gluon connects to the final hard gluon line, those which are suppressed at large $N_c$ are barred in Fig.~\ref{fig:num}. Note that one diagram contributing to the hard process in Fig.~\ref{fig:den} can also be dropped at large $N_c$.
\ei


Applying the pictorial rules of Fig.~\ref{fig:pictorial-rules} we find ($\Kvec = \Kvec_1$)
\bea
H &=& \frac{N_c^3}{4} \frac{\qvec^2}{\Kvec^2 (\Kvec-\qvec)^2} \left[1 + \frac{(1-\xh)^2 \Kvec^2}{(\Kvec-\xh \qvec)^2} \right] \, , \label{H} \\ 
C &=&  \frac{N_c^5}{4}  \frac{\qvec^2}{\Kvec^2 (\Kvec-\qvec)^2} \, \left[ \frac{\kvec -  \ellvec}{(\kvec - \ellvec)^{2}} -  \frac{\kvec}{\kvec^{2}} \right] \cdot \frac{-(\kvec - \frac{x}{\xh} \Kvec)}{(\kvec - \frac{x}{\xh}\Kvec)^{2}}  \, . \label{C}
\eea
Inserting the latter expressions in \eq{specn1-qg-0}, we find that to logarithmic accuracy, the spectrum arises from the $k_\perp$-domain $x |\Kvec| \ll |\kvec| \ll |\ellvec| \sim \mu$ (recall that $\xh \sim \morder{1}$) and reads
\be
\label{specn1-qg-LL}
\left. x \frac{\dd I^{(1)}}{\dd x}\right|_{q\to qg}  =  \left[ 1+\frac{(1-\xh)^2 \Kvec^2}{(\Kvec-\xh \qvec)^2} \right]^{-1} \, \frac{N_c \alpha_s}{\pi} \, \frac{L}{\lambda_g} \, \log{\left(\frac{\mu^2 }{x^2 \Kvec^2} \right) } \, .
\ee
The calculation to all orders in opacity can be done as for $1 \to 1$ processes \cite{Peigne:2014uha}. Quite intuitively, as was the case for $1 \to 1$ (see comments after \eq{specn1-LL}), the result is formally obtained by replacing $L/\lambda_g \to 1$ and $\mu^2 \to \Delta q_\perp^2(L)$ in the first order result \eq{specn1-qg-LL},
\bea
\label{qqg-spec}
\left. x \frac{\dd I}{\dd x}\right|_{q\to qg}  &=& \sum_{n=1}^{\infty} \left. x \frac{\dd I^{(n)}}{\dd x}\right|_{q\to qg} = \kappa_{q \to qg}  \, \frac{N_c \alpha_s}{\pi} \, \log{\left(\frac{\Delta q_\perp^2(L)}{x^2 \Kvec^2} \right) }   \, , \\
\label{kappa}
&& \kappa_{q \to qg} \equiv \frac{ (\Kvec-\xh \qvec)^2}{(\Kvec-\xh \qvec)^2 + (1-\xh)^2 \Kvec^2}  \, .
\eea

To logarithmic accuracy, the radiation spectra associated to the $q\to qg$ and $g \to g$ hard processes, given in \eq{qqg-spec} and \eq{spec-alln-LL}, are proportional to the same logarithm (up to the renaming of the hard scale $\qvec \to \Kvec$ in the logarithm of \eq{spec-alln-LL}), and otherwise differ by the overall factor $\kappa_{q \to qg}$ depending on the kinematical variables defining the $q \to qg$ hard process. In the kinematical situation where $|\qvec| \ll |\Kvec|$ and $\xh = 1/2$, we recover the factor $\kappa_{q \to qg} = 4/5$ found in Ref.~\cite{Liou:2014rha}. 

Finally, we stress that the result \eq{qqg-spec} arising from the region 
\be
\label{krange-qqg}
x K_\perp \ll k_\perp \ll \Delta q_\perp 
\ee 
was obtained using the assumption \eq{tf-assumption}. Our derivation of \eq{qqg-spec} is thus strictly valid provided the condition $\tf \gg t_{\rm hard}$, or equivalently $x \gg k_\perp^2/K_\perp^2$, holds in the whole domain \eq{krange-qqg}. This implies the following validity range of \eq{qqg-spec},
\be
\label{validity-range}
\frac{\Delta q_\perp^2}{K_\perp^2} \, \ll \, x \, \ll \, \frac{\Delta q_\perp}{K_\perp}  \, (\ll 1) \, .
\ee

Let us remark that in Ref.~\cite{Liou:2014rha} the calculation of the spectrum associated to $q \to qg$ (done using a different kinematics, namely, small $|\qvec|$ and $\xh = 1/2$ in our notations) does not assume $\tf \gg t_{\rm hard}$. Its range of validity is thus broader than the range \eq{validity-range} and extends to $x$-values which are smaller than the lower bound in \eq{validity-range}. It is likely that the range of validity of the spectrum \eq{qqg-spec} similarly extends beyond \eq{validity-range}, for any $|\qvec|$ and $\xh$. However keeping track of contributions with $\tf \lsim t_{\rm hard}$ would greatly complicate our calculation.

\subsection{Interpretation and conjecture}

Here we give a simple interpretation of the factor $\kappa_{q \to qg}$ (given in \eq{kappa}), as well as of the color factor $N_c$ appearing in front of the logarithm in \eq{qqg-spec}.

For $N_c \geq 3$, the final quark-gluon pair produced in the hard $q \to qg$ process can be in three different irreducible color representations,
\be
\label{quark-glu}
{\bf 3 \otimes 8 = 3 \oplus \bar{6} \oplus 15} \, ,
\ee
where the names of the representations indicate their dimensions in the particular case $N_c =3$. For general $ N_c$ the three representations have dimensions
\be
K_3= N_c, \ \ \  K_6=  \frac{N_c (N_c -2) (N_c+1)}{2}, \ \ \ K_{15}= \frac{N_c (N_c +2) (N_c-1)}{2}\, ,
\ee
and Casimir operators
\be
C_3= \frac{N_c^2-1}{2 N_c}, \ \ \  C_6=  \frac{(N_c -1) (3 N_c+1)}{2N_c}, \ \ \ C_{15}= \frac{(N_c +1) (3 N_c-1)}{2N_c}\, .
\ee

We observe that the diagrams of Fig.~\ref{fig:num} which are suppressed in the large $N_c$ limit are those where the final $qg$ pair is produced in the fundamental representation ${\bf 3}$. This is not surprising, since we have seen in section \ref{sec:gtog} that the coherent radiation associated to the $q \to q$ process is suppressed at large $N_c$ (see footnote~\ref{foot:quark-loss}). Thus, in Fig.~\ref{fig:num} (and in Fig.~\ref{fig:num} only, see the comments below) we may remove from the beginning the `triplet' component of the final $qg$ pair (or equivalently, project the latter on the ${\bf \bar{6} \oplus 15}$ subspace). At large $N_c$, this is simply achieved by replacing $T^bT^a \to 0$ in \eq{GB}, leaving $T^aT^b$ unchanged. The hard production amplitude with the final quark-gluon triplet component {\it removed} thus reads
\be
\label{GB-triplet-removed}
{\cal M}_{\rm hard}^{{\bf \bar{6} \oplus 15}} \  \propto  \ \ T^aT^b \left( \frac{\Kvec}{\Kvec^{2}} - \frac{\Kvec - \qvec}{(\Kvec - \qvec)^{2}} \right)  \ \, .
\ee
Squaring this we find 
\be
|{\cal M}_{\rm hard}^{{\bf \bar{6} \oplus 15}}|^2 =  \frac{N_c^3}{4} \frac{\qvec^2}{\Kvec^2 (\Kvec-\qvec)^2} \, ,
\ee
and dividing by the expression \eq{H} we get
\be
\label{color-select}
\frac{|{\cal M}_{\rm hard}^{{\bf \bar{6} \oplus 15}}|^2}{|{\cal M}_{\rm hard}|^2} = \frac{ (\Kvec-\xh \qvec)^2}{(\Kvec-\xh \qvec)^2 + (1-\xh)^2 \Kvec^2} = \kappa_{q \to qg}  \, .
\ee
Thus, the factor $\kappa_{q \to qg}$ is interpreted as the probability that the quark-gluon pair produced in $q \to qg$ is {\it not} in the `triplet' color representation. 

Thus, the dependence of the spectrum \eq{qqg-spec} on the hard process kinematical variables $\qvec$, $\Kvec$, $\xh$, arises from the constraint that at large $N_c$, only non-triplet $qg$ pairs can contribute to the set $C$ of diagrams (Fig.~\ref{fig:num}). This `selection' of the ${\bf \bar{6} \oplus 15}$ subspace is due to the specific connection of the soft radiated gluon between initial and final state in Fig.~\ref{fig:num}. In particular, it would be incorrect to attribute this effect to the smaller dimension $K_3 = N_c$ of the triplet representation as compared to the dimension of the ${\bf \bar{6} \oplus 15}$ subspace ($K_6+K_{15}=N_c^3$ at large $N_c$). For instance, the lower dimension of ${\bf 3}$ does not prevent the $qg$ pair to be produced as a triplet in the hard $q \to qg$ process (see set $H$ of diagrams, Fig.~\ref{fig:den}), even at large $N_c$.  

The logarithmic range \eq{krange-qqg} can be interpreted in a similar way as the range \eq{krange-zeromass} for $1\to 1$ processes (see section~\ref{interpret-11}). Moreover, in the present $q \to qg$ case, the condition $x K_\perp \ll k_\perp$ written as $1/k_\perp \gg \Delta r_\perp \sim v_\perp \,\tf \sim (K_\perp/E) \cdot (\omega/k_\perp^2)$ (similarly to \eq{interpret-krange-zeromass}) means that at the time of its emission, the radiated gluon does not probe the transverse size $\Delta r_\perp$ of the $qg$ pair. From the point of view of soft radiation, the $qg$ pair thus behaves as an effectively pointlike system. 

Finally, the color factor $N_c$ in \eq{qqg-spec} can be simply understood from the rule \eq{color-rule}. Indeed, since coherent radiation arises from a kinematical domain where the $qg$ pair is effectively pointlike, the result should depend on its {\it total} color charge, not on the color of its separate constituents. Since the color state of the final $qg$ pair in Fig.~\ref{fig:num} is either ${\bf \bar{6}}$ or ${\bf 15}$, and these two representations have the same Casimir operator at large $N_c$, namely $3N_c/2$, the rule \eq{color-rule} gives ($R' = {\bf \bar{6}}$ or ${\bf 15}$)
\be
\label{color-rule-final}
2 T_{\bf 3}^a T_{R'}^a = C_{\bf 3} + C_{R'} - C_{\bf 8} = \frac{N_c}{2} +  \frac{3 N_c}{2} - N_c = N_c \, .
\ee

\subsubsection*{A conjecture}

Guided by the above interpretation of our result, we conjecture the following simple formula for the medium-induced radiation spectrum associated to hard forward $1 \to n$ processes (where the $n$ final-state partons have finite longitudinal momentum fractions $x_i = K_i^+/p^+ \sim \morder{1}$ and transverse momenta $\Kvec_i$ of similar magnitude $\sim |\Kvec|$), 
\be
\label{conjecture}
\left. x \frac{\dd I}{\dd x}\right|_{1\to n}  =   \left[ \sum_{R'} P_{R'} (C_R + C_{R'} - C_t) \right] \frac{\alpha_s}{\pi} \, \log{\left(\frac{\Delta q_\perp^2(L)}{x^2 \Kvec^2} \right) }  \, ,
\ee
with $C_R$ and $C_t$ the color charges of the incoming parton and of the $t$-channel exchange, and $P_{R'}$ the probability for the (effectively pointlike) $n$-parton state to be produced in the color representation $R'$ in the hard process. ($P_{R'}$ may depend on the kinematical variables $x_i$, $\Kvec_i$ defining the hard $1 \to n$ process, as in the $q \to qg$ case.) 

As a first illustration, the spectrum associated to $g \to q \bar{q}$ derived in Ref.~\cite{Liou:2014rha} can be obtained from \eq{conjecture} by setting $P_{R'} = P_{\bf 8} =1$ (at large $N_c$ the final $q \bar{q}$ is color octet with unit probability) and $C_R = C_{R'} = C_t = N_c$. Not surprisingly, since the final $q \bar{q}$ is effectively pointlike, the result is the same as for the $g \to g$ process considered in Refs.~\cite{Arleo:2010rb,Peigne:2014uha}. 

As a second example, let us mention that we explicitly verified \eq{conjecture} in the case of the $g \to gg$ process. Using the same theoretical setup (including the large $N_c$ limit) and following the same procedure as for the $q \to qg$ process, we find an expression for the radiation spectrum similar to \eq{qqg-spec}, but with a different overall factor $\kappa$, 
\bea
\label{ggg-spec}
&& \hskip 6mm \left. x \frac{\dd I}{\dd x}\right|_{g\to gg}  = \kappa_{g \to gg}  \, \frac{N_c \alpha_s}{\pi} \, \log{\left(\frac{\Delta q_\perp^2(L)}{x^2 \Kvec^2} \right) }   \, , \\
\label{kappaggg}
&& \kappa_{g \to gg} \equiv  1+ \frac{(\Kvec-\xh \qvec)^2}{(\Kvec-\xh \qvec)^2 + \xh^2 (\Kvec-\qvec)^2 + (1-\xh)^2 \Kvec^2}  \, .
\eea
(Note that $\kappa_{g \to gg} = 5/3$ when $|\qvec| \ll |\Kvec|$ and $\xh = 1/2$.) To check whether \eq{ggg-spec}, \eq{kappaggg} coincide with \eq{conjecture}, we must sum in \eq{conjecture} over the different representations $R'$ of the final $gg$ pair. At large $N_c$, a two-gluon system can be in six color representations \cite{Dokshitzer:2005ek}, 
\be
\label{gluglu}
{\bf 8 \otimes 8 = 8_a \oplus 10 \oplus 1 \oplus 8_s \oplus 27 \oplus 0} \, ,
\ee
where as in \eq{quark-glu} the representations are labelled according to their dimensions in the case $N_c =3$. In particular ${\bf 0}$ is a symmetric representation which is absent when $N_c =3$. For $N_c >3$ the representations appearing in the r.h.s. of \eq{gluglu} have the Casimir operators $N_c$, $2N_c$, $0$, $N_c$, $2(N_c+1)$ and $2(N_c-1)$, respectively~\cite{Dokshitzer:2005ek}. Thus, at large $N_c$ the bracket in \eq{conjecture} reads
\be
\label{sum-rep}
\sum_{R'} P_{R'} (C_R + C_{R'} - C_t) = \sum_{R'} P_{R'} \, C_{R'} = (2 - P_{\bf 8_{\rm \bf a}} - P_{\bf 8_{\rm \bf s}})\,N_c \, ,
\ee
where we used $C_R = C_t = N_c$, probability conservation $\sum_{R'} P_{R'} = 1$, and the fact that the probability $P_{\bf 1}$ for the final $gg$ pair to be color singlet is suppressed at large $N_c$. The probability $P_{\bf 8} \equiv P_{\bf 8_{\rm \bf a}} + P_{\bf 8_{\rm \bf s}}$ to produce a {\it color octet} $gg$ pair can be simply evaluated using pictorial rules for the projection operators on specific color representations~\cite{Dokshitzer:2005ek}. Analogously to what was done in \eq{color-select} the calculation gives
\be
\label{P8}
P_{\bf 8} = \frac{|{\cal M}_{g \to gg}^{{\bf 8_{\rm \bf a}}}|^2 + |{\cal M}_{g \to gg}^{{\bf 8_{\rm \bf s}}}|^2}{|{\cal M}_{g \to gg}|^2} = \frac{\xh^2 (\Kvec-\qvec)^2 + (1-\xh)^2 \Kvec^2}{(\Kvec-\xh \qvec)^2 + \xh^2 (\Kvec-\qvec)^2 + (1-\xh)^2 \Kvec^2} \, .
\ee
Plugging \eq{P8} in \eq{sum-rep}, we see that the expression \eq{conjecture} reproduces Eqs.~\eq{ggg-spec}, \eq{kappaggg}. This completes the check of the conjectured expression \eq{conjecture} in the case of the $g \to gg$ process. 

\acknowledgments

S.P. would like to thank Tseh Liou and Al Mueller for a rich and instructive correspondence, which motivated the present study. We also thank Fran\c{c}ois Arleo for useful discussions and comments on the manuscript. Feynman diagrams have been drawn with the JaxoDraw software~\cite{Binosi:2008ig}. This work is funded by ``Agence Nationale de la Recherche'' under grant ANR-PARTONPROP.


\providecommand{\href}[2]{#2}\begingroup\raggedright\endgroup

\end{document}